# Non-Hermitian-induced higher-order topological phases in acoustic fractal lattices

Shuanghuizhi Li [1, 2, ‡], Bowei Wu [1, 2, ‡], Tingfeng Ma [1, 2, *], Haixiao Wang [3], Jiaqi Zhang [1, 2], Chenbowen Lou [1, 2]

[1] Zhejiang-Italy Joint Lab for Smart Materials and Advanced Structures, School of Mechanics and Engineering Science, Ningbo University, Ningbo 315211, China.

[2] Key Laboratory of Impact and Safety Engineering, Ministry of Education, Ningbo University, Ningbo 315211, China

[3] School of Physical Science and Technology, Ningbo University, Ningbo 315211, China.

‡ These authors contributed equally to this work.

* Author to whom correspondence should be addressed: matingfeng@nbu.edu.cn

## Abstract

For fractal (non-integer-dimension) geometries in Hermitian system, high-order topological phases were realized in the past few years, for which, it is difficult to tune the energy concentration degree of topological states flexibly and smoothly. Non-Hermiticity method is expected to solve this problem. However, in non-integer-dimension system, the non-Hermitian-fractal coupled topological mechanism is unclear and topological phases induced by non-Hermiticity remain unrealized. By introducing a loss contrast in a fractal lattice, this study proposes a non-Hermitian route to realize higher-order topological phases in acoustic fractal lattices with good tunability. Based on the tight-binding approximation, calculations on the Hamiltonian of the system yield the wave-function distribution of the zero-energy modes, revealing the formation mechanisms and conditions for the topological phase transitions induced by non-Hermiticity in acoustic fractal lattices. We numerically and experimentally realize non-Hermitian-induced topological edge and corner states in a fractal structure. Furthermore, a tunable acoustic energy concentrator has been realized, namely, the degree of acoustic energy localization can be tuned conveniently by merely adjusting the loss contrast, rather than redesigning the structure parameters. This work not only establishes an effective mechanism for manipulating higher-order topology in complex fractal geometries through non-Hermiticity but also provides a theoretical framework for exploring exotic topological states of matter in non-integer dimensions.

## I. Introduction

In recent years, topological concepts have provided powerful approaches for manipulating classical waves beyond conventional mechanisms [1]. By exploiting nontrivial band topology and the associated topological edge states, robust wave transport can be achieved against certain classes of defects and disorders, as demonstrated in a variety of acoustic topological systems [2–4]. As an important branch of this field, topological acoustics has enabled unconventional phenomena through phononic crystals and acoustic metamaterials, including robust edge transport and topological localization, thereby offering new opportunities for acoustic filtering, directional wave guiding, and sensing. More recently, the emergence of higher-order topological insulators (HOTIs) has extended the conventional bulk-edge correspondence to higher codimensions [5–7]. In an $n$-th-order topological insulator of dimension $d$, topologically protected edge states can emerge in $(d-n)$-dimensional subspaces, with zero-dimensional corner states in two-dimensional systems being a representative example. Higher-order topology has subsequently been experimentally realized in various acoustic and mechanical systems, including quadrupole, Kagome, and sonic-crystal configurations [8–11]. Recent developments have further extended acoustic higher-order topology to three-dimensional multipole phases, nodal-line phases, topological transport, and non-Abelian multiband systems [12–15], demonstrating increasingly diverse mechanisms for localizing and manipulating acoustic waves at lower-dimensional boundaries.

Most early studies of topological acoustics were formulated within the Hermitian framework, in which intrinsic dissipation was commonly neglected or treated as an undesirable perturbation. In realistic acoustic systems, however, radiation, viscous and thermal losses are generally unavoidable, naturally motivating the extension of topological physics to non-Hermitian systems [16,17]. Non-Hermitian systems exhibit complex spectra and nonorthogonal eigenstates and possess topological classifications fundamentally different from their Hermitian counterparts [18,19]. Recent advances in non-Hermitian elastodynamics have extended these concepts to active mechanical metamaterials, where nonreciprocal couplings enable unconventional wave localization and topological mode manipulation [20, 21]. Programmable mechanical platforms further provide flexible control of non-Hermitian topological transitions and exceptional degeneracies [22, 23]. Representative phenomena include parity-time-symmetry breaking and exceptional degeneracies [24,25], as well as the non-Hermitian skin effect and the associated breakdown of conventional Bloch bulk-edge correspondence [26–28]. Non-Hermiticity can also strongly reshape conventional topological edge localization [29]. Its interplay with higher-order topology has further led to second-order non-Hermitian phases and higher-order skin effects [26–29]. More importantly, non-Hermiticity can serve as an additional degree of freedom for topological engineering: spatially patterned losses can reconfigure topological modes [34], while deliberately introduced losses can directly induce higher-order topological phases in acoustic crystals [35]. Recent studies further demonstrate that loss can generate effective nonreciprocity and hybrid-order skin localization [36], and can even induce higher-order topology solely through the embedding of lossy sites [37]. These results establish loss as an active control parameter rather than merely a detrimental perturbation.

Fractal structures, characterized by self-similar hierarchy and non-integer dimensionality, provide another intriguing platform for exploring unconventional wave physics [38–40]. Their multiscale geometries and abundant internal boundaries enable distinctive spectral, propagation, and localization behaviors beyond those of conventional periodic lattices. Higher-order topological phases have been theoretically established in Sierpiński fractal lattices and subsequently demonstrated experimentally in acoustic fractal systems, where abundant inner corner, outer corner and edge states emerge from the hierarchical geometry [41–43]. More recently, spin-Chern topology has been experimentally realized in a phononic fractal lattice [44], while fractality-induced topological phase squeezing and general real-space construction principles have further broadened the topological landscape of fractal systems [45,46]. Nevertheless, existing acoustic fractal topological systems are predominantly controlled through geometric parameters or coupling configurations, which limits their post-fabrication tunability. Meanwhile, non-Hermitian physics has increasingly begun to intersect with fractal systems, including inner non-Hermitian skin effects [47], self-similar non-Hermitian quantum-fractal spectra [48], and most recently, fractal braiding of non-Hermitian complex-energy bands [49]. However, whether spatially engineered dissipation can directly drive higher-order topological phase transitions in acoustic fractal lattices and simultaneously provide controllable manipulation of their corner and edge states remains largely unexplored.

In this work, we investigate loss-induced higher-order topology in a non-Hermitian acoustic fractal lattice. Based on the tight-binding approximation, we construct a non-Hermitian Hamiltonian for the fractal lattice and systematically reveal the evolution of its higher-order topological phases with the increasing loss contrast. The corresponding topological phase transitions and the emergence of zero-energy corner and edge states are characterized theoretically and further verified experimentally. By combining the eigenstate distributions with the local density of states, we establish the relationship between the spatial distribution of dissipation and the localization characteristics of the resulting topological states. Furthermore, we demonstrate loss-tunable acoustic energy localizations, namely, the degree and spatial distribution of acoustic localization can be conveniently controlled by varying the loss contrast without modifying the underlying fractal geometry. These results establish dissipation as an effective tuning degree of freedom for higher-order topological states in acoustic fractal systems and provide a route toward reconfigurable wave localization in complex topological structures.

## II. Construction and topological phase calculation of the non-Hermitian acoustic fractal system

Fig. 1 illustrates the construction of the non-Hermitian acoustic Sierpiński fractal lattice. The fractal geometry is generated by recursively removing the central region of a square lattice, yielding a Hausdorff dimension $d_f = ln8/ln3$. In the acoustic implementation, each lattice site is represented by an acoustic resonator coupled to its neighboring resonators. The first iteration defines the G1 fractal, while successive recursive removals of the central regions generate the G2 and higher-generation structures, as shown in Fig. 1(a).

To realize the local coupling configuration required for quadrupole topology, we employ a 16-resonator building block, as illustrated in Fig. 1(b). Following the π-flux configuration underlying the BBH quadrupole model [5, 6], nearest-neighbor couplings are assigned equal magnitudes but opposite signs, with a common coupling strength. Unlike the conventional Hermitian BBH model, where the topological transition is driven by dimerization between intra- and intercell coupling amplitudes, the present scheme retains uniform coupling magnitudes and instead establishes the required π-flux pattern through the spatial arrangement of positive and negative couplings, while non-Hermiticity is introduced by a spatially inhomogeneous loss distribution. In the acoustic implementation, the opposite coupling signs are realized by shifting the relative position of the connecting waveguide between neighboring resonators. The corresponding eigenmodes of a coupled-resonator pair are shown in Fig. 1(c), and the resulting acoustic realization is presented in Fig. 1(d).

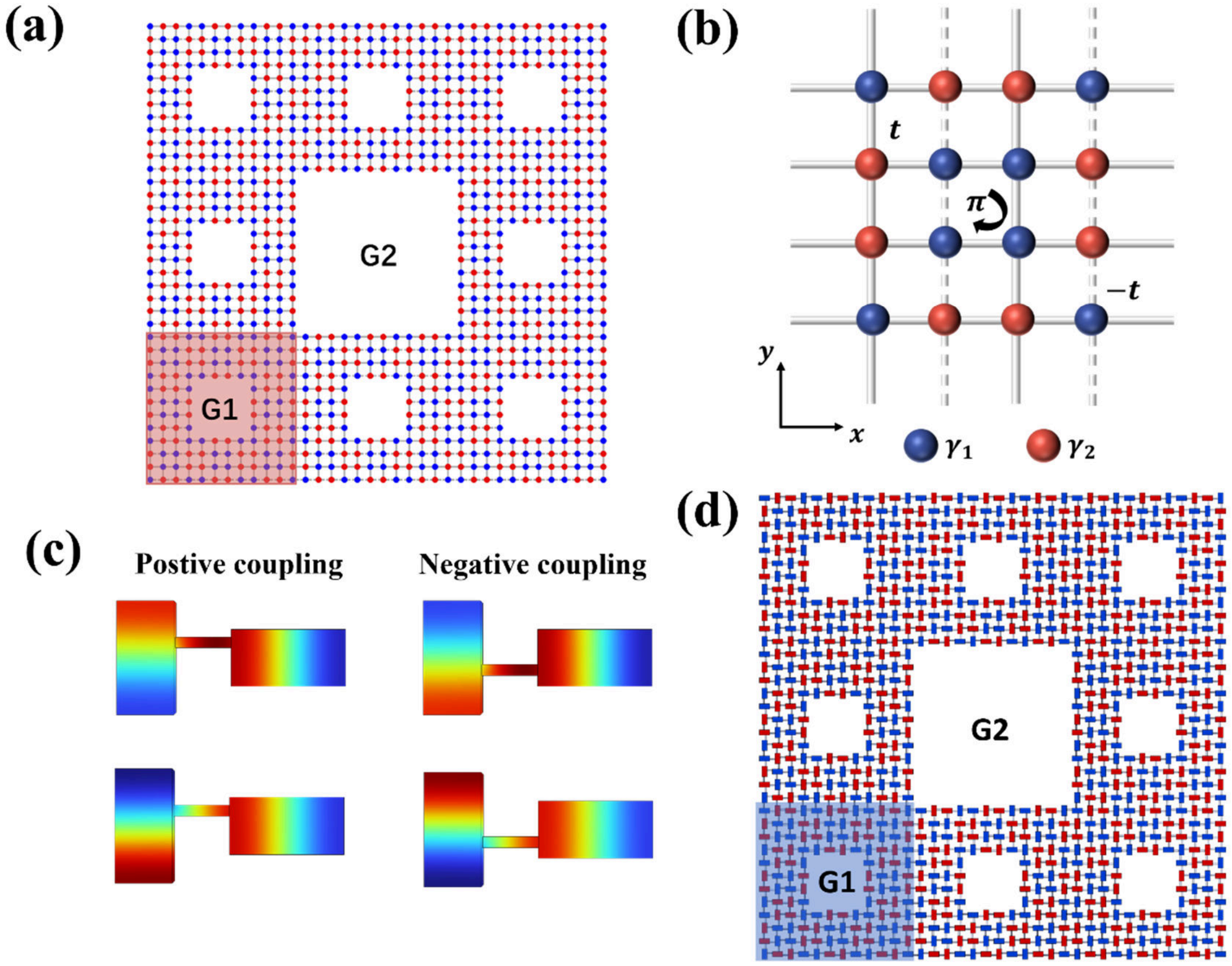

**FIG. 1.** Construction of the non-Hermitian acoustic Sierpiński fractal lattice. (a) Schematic illustration of the G1 and G2 Sierpiński fractal lattices. (b) Sixteen-site building block of the tight-binding model. Solid and dashed bonds denote positive and negative nearest-neighbor couplings, respectively, with equal coupling magnitude $t$. Blue and red sites represent resonators with background loss $\gamma_{bg}$ and enhanced loss $\gamma_{loss}$, respectively. The spatially patterned loss contrast is defined as $\Delta\gamma = \gamma_{loss} - \gamma_{bg}$. (c) Representative eigenmode distributions of two coupled acoustic resonators for the positive- and negative-coupling configurations, illustrating the realization of opposite coupling signs by changing the relative position of the connecting waveguide. (d) Acoustic implementation of the fractal lattice, where blue and red indicate resonators with background and enhanced losses, respectively.

The system comprises two classes of resonators. The background sites are subject only to an intrinsic loss $\gamma_{bg}$, whereas the designated lossy sites experience an enhanced loss $\gamma_{loss}$. The strength of the spatially inhomogeneous dissipation is therefore characterized by the loss contrast: $\Delta\gamma = \gamma_{loss} - \gamma_{bg}$. Accordingly, the relevant control parameter governing the topological spectral reconstruction is the nonuniform loss contrast $\Delta\gamma$, rather than the absolute magnitude of the background loss. This mechanism is consistent with previously reported loss-induced acoustic quadrupole topological systems, in which the contrast between the imaginary on-site potentials of the two sublattices provides the key ingredient for opening a topological band gap [31].

In this system, the finite non-Hermitian fractal lattice under open edge conditions is described by the tight-binding Hamiltonian:

$$H = \sum_{\langle m,n\rangle} t_{mn}\left(\hat{c}_m^\dagger \hat{c}_n + \hat{c}_n^\dagger \hat{c}_m\right) - i\sum_n \gamma_n \hat{c}_n^\dagger \hat{c}_m \tag{1}$$

where $t_{mn}$ denotes the positive or negative nearest-neighbor coupling associated with the $\pi$-flux configuration, and $\hat{c}_n^\dagger$ $(\hat{c}_n)$ is the creation (annihilation) operator at the $n_{th}$ lattice site. $\gamma_n$ denotes the site-dependent loss, with $\gamma_{bg}$ representing the background loss and $\Delta\gamma = \gamma_{loss} - \gamma_{bg}$ representing the additional loss contrast. The detailed derivations are shown in Appendix C.

Because translational symmetry is absent in the fractal lattice, conventional momentum-space topological invariants formulated in terms of Bloch wave vectors are not directly applicable. We therefore characterize the topology of the finite non-Hermitian fractal system using a biorthogonal real-space quadrupole moment. For a non-Hermitian Hamiltonian, the left and right eigenstates are generally distinct; accordingly, the real-space quadrupole invariant must be constructed within the corresponding biorthogonal spectral subspaces:

$$Q_{xy} = \frac{1}{2\pi}\mathrm{Im}\left\{\log\left[\det\left(U_L^\dagger \hat{Q} U_R\right)\sqrt{\det\left(\hat{Q}^\dagger\right)}\right]\right\} \tag{2}$$

where $\hat{Q} = \exp\left[i2\pi\hat{x}\hat{y}/\left(N_x N_y\right)\right]$ is the real-space quadrupole operator. $U_L$ and $U_R$ are constructed from the occupied left and right eigenstates with Re(En) < 0, respectively, and satisfy the biorthonormal condition $U_L^\dagger U_R = I$. Here, $\hat{x}$ and $\hat{y}$ denote the position operators along the $x$ and $y$ directions, respectively, while $N_x$ and $N_y$ denote the corresponding linear dimensions of the complete lattice.

Fig. 2 shows the calculated biorthogonal real-space quadrupole moment $Q_{xy}^{NH}$ as a function of the normalized loss contrast $\Delta\gamma/t$, for the G1, G2, and G3 fractal lattices. At weak loss contrast, all three structures remain in the trivial regime with $Q_{xy}^{NH} = 0$. As $\Delta\gamma/t$ increases, the quadrupole moment undergoes a sharp transition from 0 to the quantized value 0.5. For the finite lattices considered here, the transition occurs at approximately $\Delta\gamma/t$ =2.575, 2.375 and 2.325 for G1, G2, and G3, respectively. Thus, all three fractal generations exhibit the same loss-driven topological transition, while the corresponding finite-size transition thresholds show a moderate generation dependence.

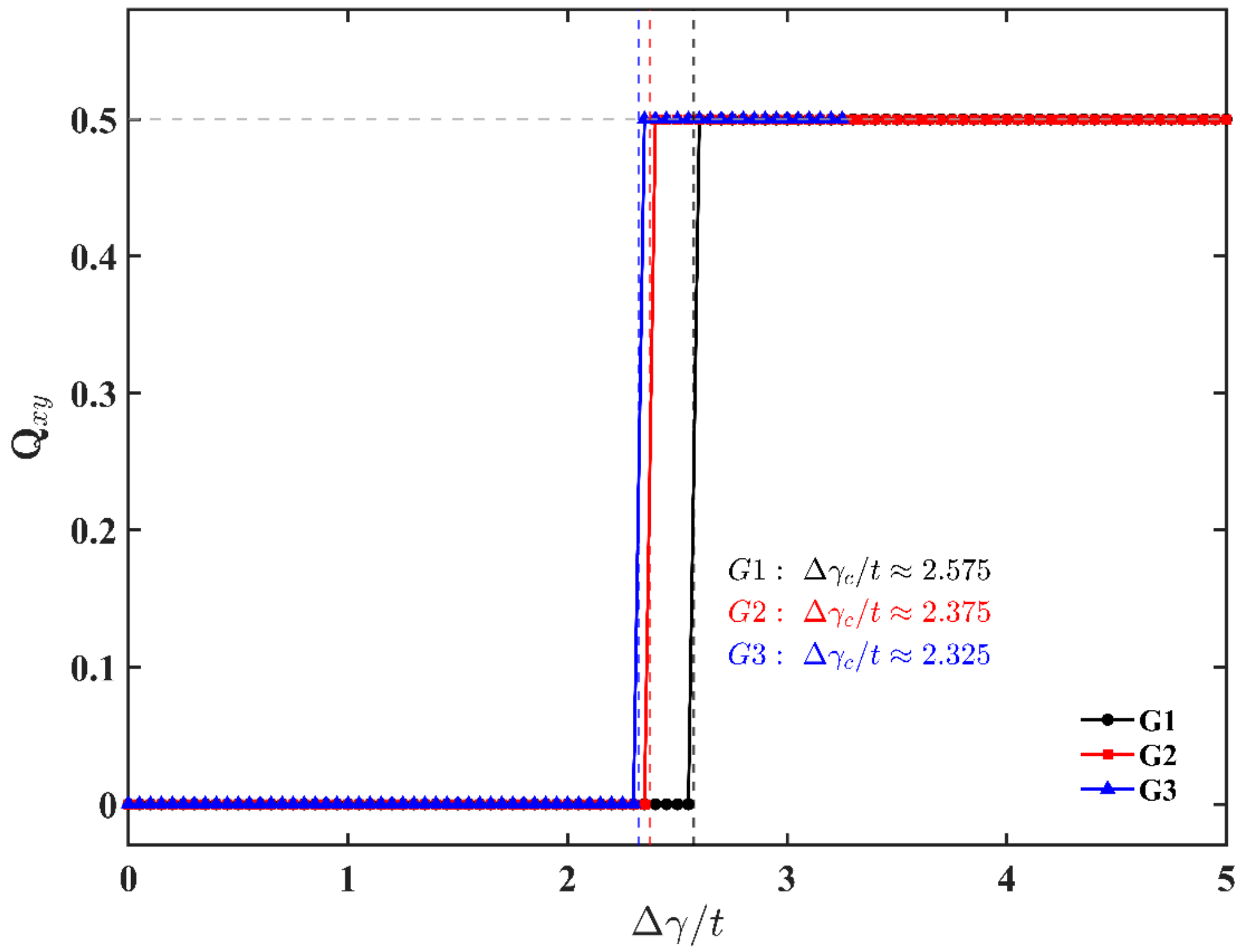

**FIG. 2.** Loss-induced topological phase transition characterized by the biorthogonal real-space quadrupole moment. The non-Hermitian real-space quadrupole moment $Q_{xy}^{NH}$ is plotted as a function of the normalized loss contrast $\Delta\gamma/t$, for the finite G1, G2, and G3 Sierpiński fractal lattices under open edge conditions. The transition from $Q_{xy}^{NH} = 0$ to the quantized value $Q_{xy}^{NH} = 0.5$ identifies the loss-induced transition from the trivial to the nontrivial quadrupole phase. For the present finite structures, the transition points are approximately $\Delta\gamma/t = 2.575,\ 2.375,\ 2.325$ and for G1, G2, and G3, respectively.

Now the loss-induced topological transition has been realized through the biorthogonal real-space quadrupole moment $Q_{xy}^{NH}$, we next examine whether the resulting hierarchical topological states can be realized in realistic acoustic structures. Full-wave finite-element analyses are performed for finite G1 and G2 acoustic Sierpiński fractal lattices. Fig. 3(a) shows the real parts of the complex eigenfrequencies of the G1 structure in the nontrivial regime. Based on their spatial profiles, the eigenmodes can be classified into bulk, outer-edge, outer-corner, and inner-corner states. The corner states reside within the real-frequency gap and are spectrally well separated from the extended bulk states. The corresponding full-wave pressure-field distributions are shown in Figs. 3(c)–(e). The outer-corner states are predominantly localized at the outermost corners of the G1 fractal, whereas the inner-corner states are concentrated around the corners of the central fractal cavity. In contrast, the outer-edge states extend primarily along the external edge. The coexistence of outer- and inner-corner states originates from the intrinsic internal boundaries of the Sierpiński geometry. Unlike conventional quadrupole topological lattices without internal cavities, the recursive fractal geometry generates additional internal boundaries and associated corners, thereby creating additional localization sites for the topological modes and giving rise to a hierarchical corner-state structure across the outer and inner boundaries.

For the G2 structure, the higher fractal hierarchy gives rise to a substantially richer set of spatially localized modes. As shown in Fig. 3(b), the increased number of recursively generated internal boundaries supports additional classes of localized eigenstates. In addition to the bulk, outer-edge, and outer-corner states, two distinct types of inner-corner states and an inner-edge state can be identified. Representative full-wave field distributions of these modes are shown in Figs. 3(f)–(k). The outer-corner state remains localized at the outermost vertices of the entire structure, whereas the two inner-corner states are concentrated at geometrically distinct corners of internal boundaries belonging to different levels of the fractal hierarchy. The inner-edge state, by contrast, is primarily confined along an internal fractal edge. Thus, increasing the fractal generation does not alter the underlying mechanism of the loss-induced topology; rather, it substantially enriches the spatial hierarchy of the topological edge states by generating additional internal boundaries.

The full-wave results in Fig. 3 are further compared with the tight-binding calculations presented in Appendix A. In particular, the outer-corner, inner-corner, outer-edge, and inner-edge states predicted by the tight-binding model are all reproduced in the full-wave simulations of the acoustic structures. This correspondence demonstrates that the tight-binding Hamiltonian captures the essential physics underlying the loss-induced hierarchical topological states, while the full-wave calculations further establish their realizability in realistic acoustic structures.

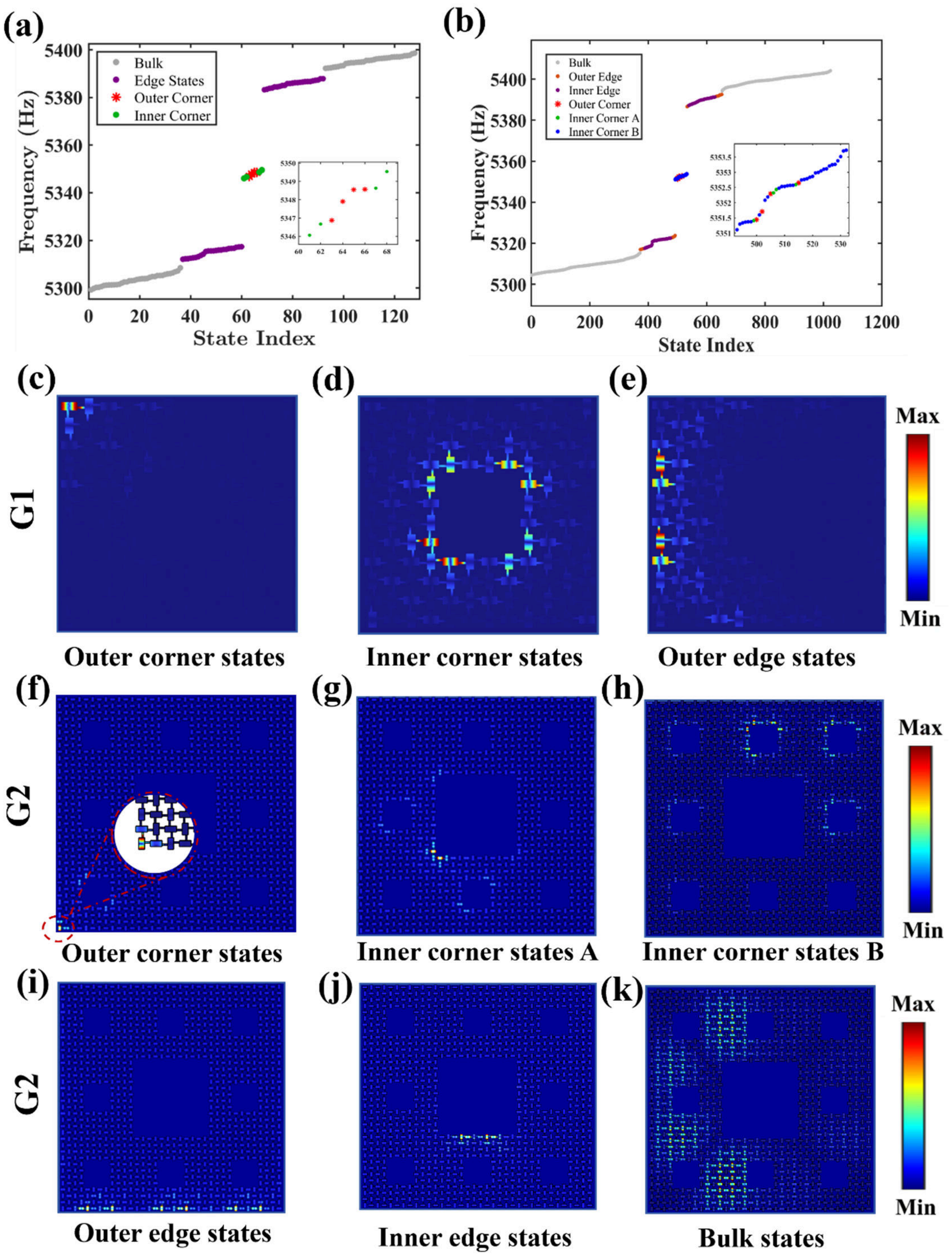

**FIG. 3.** Full-wave simulation of hierarchical topological states in non-Hermitian acoustic fractal lattices. (a) Real parts of the complex eigenfrequencies of the finite G1 acoustic fractal lattice in the nontrivial regime. The eigenmodes are classified as bulk, outer-edge, outer-corner, and inner-corner

states according to their spatial localization. (b) Corresponding eigenfrequency spectrum of the G2 acoustic fractal lattice. The increased fractal hierarchy gives rise to additional states localized at the internal boundaries. The inset shows an enlarged view of the spectral region containing the corner states. (c)–(e) Representative full-wave pressure-field distributions of the G1 outer-corner, inner-corner, and outer-edge states, respectively. (f)–(k) Representative full-wave pressure-field distributions of the G2 outer-corner, type-A inner-corner, type-B inner-corner, outer-edge, inner-edge, and bulk states, respectively.

To experimentally verify the existence of the non-Hermitian-induced fractal higher-order topological states, we designed and fabricated acoustic fractal lattices composed of acoustic resonators by using stereolithography 3D printing technology, and measured acoustic responses of the experimental sample with non-uniform loss distributions. As shown in Fig. 4(a), blue sound-absorbing materials are inserted into the small holes of the rectangular resonators to introduce additional loss. Each resonator possesses two small holes that can be opened or closed by two circular cover plates. During the experiment, acoustic waves generated by a loudspeaker are introduced into the sample through the small hole on the left side of the resonator, and a microphone is used to detect the signal through the small hole on the right side of the same resonator. This measurement is repeated for all resonators in the sample.

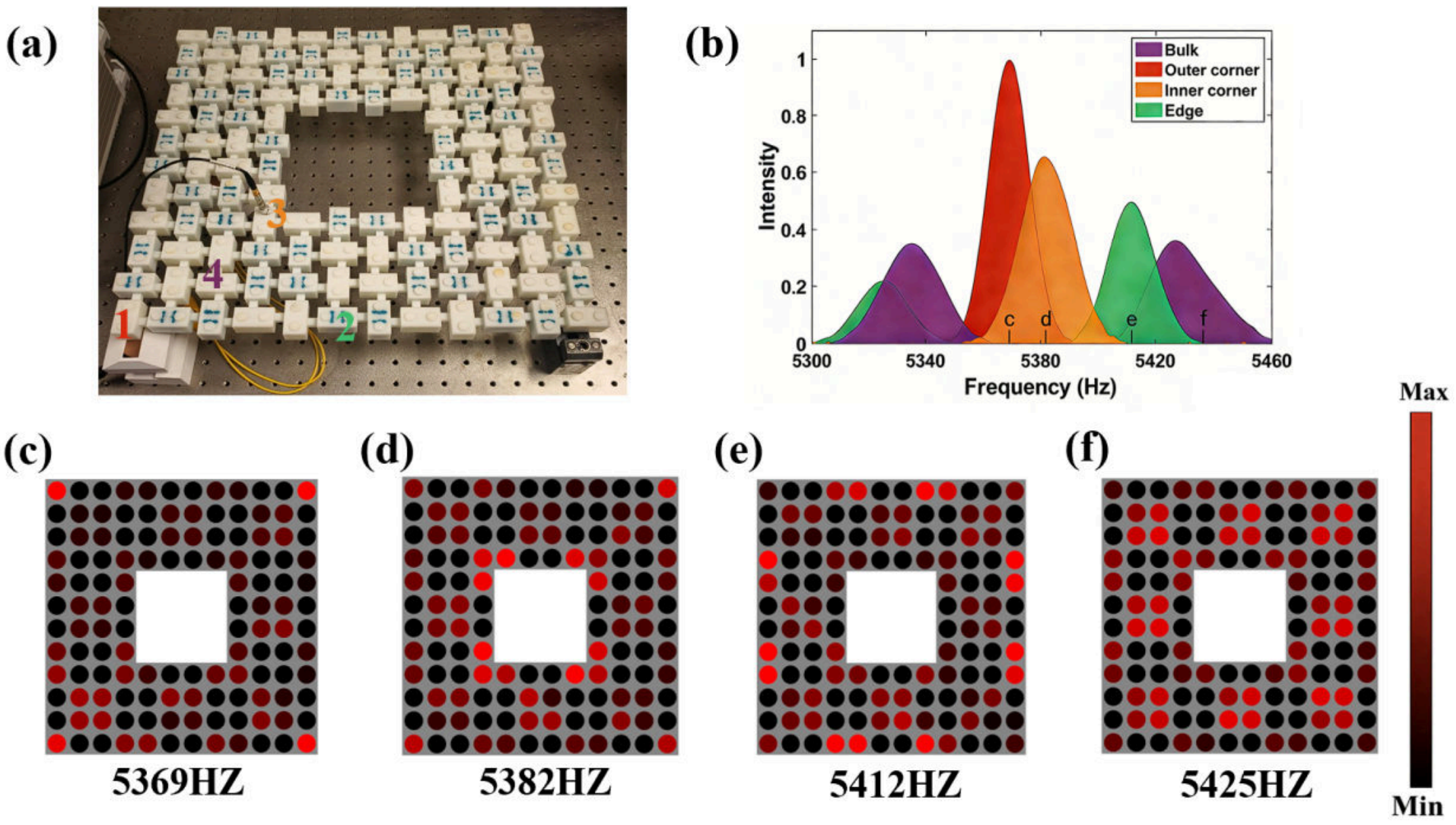

**FIG. 4.** Experimental observation of hierarchical topological states in the non-Hermitian G1 acoustic fractal lattice. (a) Photograph of the fabricated acoustic fractal sample with the prescribed

spatially nonuniform loss distribution. The marked positions 1–4 denote representative outer-corner, edge, inner-corner, and bulk resonators used for local spectral measurements. (b) Measured local response spectra at the representative sites, showing characteristic resonances associated with the different classes of modes. (c)–(f) Measured spatial acoustic-intensity distributions at 5369, 5382, 5412, and 5425 Hz, corresponding to the outer-corner, inner-corner, edge, and bulk states, respectively.

We selected the resonator located in the middle of one edge (marked as "2" in Fig. 4(a) ) and plotted its response spectrum showed by the green curves in Fig. 4(b). These two green peaks correspond to the edge states. Unlike the bulk and edge state spectra, the spectra measured at the outer and inner corner resonators (marked as "1" and "3", respectively, in Fig. 4(a) ) exhibit only a single peak near 5369 Hz and 5382 Hz, as indicated by the red and yellow curves in Fig. 4(b), which agrees well with the numerically predicted eigenfrequencies for the inner and outer corner states presented in Figs. 4(b). To further demonstrate the non-Hermitian-induced fractal higher-order topological phase, we also plotted the site-resolved responses measured at the peak frequencies of the outer corner, inner corner, edge, and bulk states spectra in Figs. 4(c)–(f), respectively. At 5369 Hz, which corresponds to the peak of the outer corner state spectrum, the acoustic intensity measured at the outer corner resonant cavities is much higher than those in other regions (Fig. 4(c)), indicating the presence of the outer corner state. In contrast, the responses measured at the peak frequencies of the inner corner state, edge state, and bulk state spectra (5382 Hz, 5412 Hz, and 5425 Hz, respectively) are higher obviously in their respective regions (Figs. 4(d)–(f)). These experimental observations agree well with the numerical simulation results presented in Fig. 2.

## Ⅲ. Flexible tuning of energy concentration of topological states based on the non-Hermitian method

In the preceding analysis, the loss contrast $\Delta\gamma$ was shown to drive the fractal lattice from a trivial phase into a nontrivial quadrupole phase characterized by $Q_{xy}^{NH} =$ 0.5. A natural question then arises as to whether, once the system enters the nontrivial regime, $\Delta\gamma$ can further serve as a continuous control parameter for tuning the spatial localization of the topological edge states. To address this question, we examine the evolution of the outer-corner, inner-corner, and edge states of the G1 fractal lattice as a function of $\Delta\gamma$.

Fig. 5(a) shows the real parts of the complex eigenfrequencies of the finite lattice as $\Delta\gamma$ is varied, with the color scales indicating the modal localization weights in the outer-corner, inner-corner, and edge regions, respectively. It is shown that after entering the nontrivial regime, the corresponding corner and edge states retain their characteristic spectral branches, while their spatial localization weights continue to evolve with increasing loss contrast. It is important to distinguish topology from localization: the former is characterized by the quantized biorthogonal real-space quadrupole moment $Q_{xy}^{NH}$, whereas the latter measures the fraction of a given eigenmode concentrated within a prescribed spatial region. Thus, even when the topological phase remains unchanged, the loss contrast can continuously reshape the spatial distribution of the topological states.

To quantify this evolution, Fig. 5(b) shows the normalized modal weights of the outer-corner, inner-corner, and edge states within their corresponding characteristic regions as functions of $\Delta\gamma$. As the loss contrast increases, all three weights increase progressively and approach unity, indicating that the eigenfields become increasingly confined to the corresponding corners or boundaries. The concurrent spectral separation and increase in localization weight further indicate that a larger loss contrast suppresses the spatial extension of these edge or corner states into the surrounding lattice, thereby enhancing their characteristic localization.

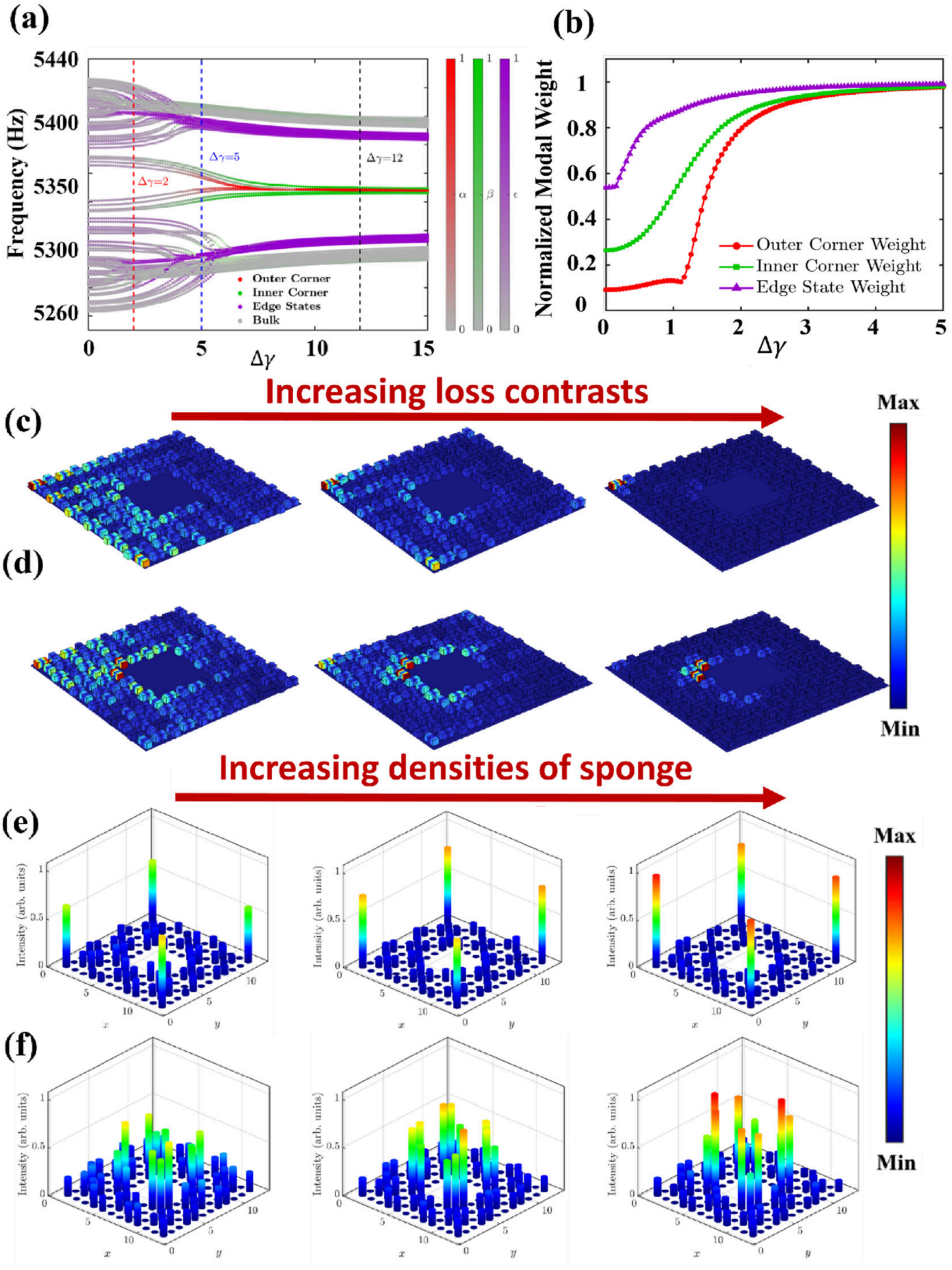

**FIG. 5.** Loss-controlled localization of topological states in the non-Hermitian acoustic fractal lattice (Tunable acoustic energy concentrator). (a) Evolution of the real parts of the complex eigenfrequencies with the increasing loss contrast $\Delta\gamma$. The color scales characterize the spatial localization weights of the outer-corner, inner-corner, and edge states. The vertical dashed lines mark representative loss contrasts selected for comparison of the modal profiles. (b) Normalized modal weights within the characteristic outer-corner, inner-corner, and edge regions as functions of $\Delta\gamma$. (c) and (d) Full-wave field distributions of representative outer-corner and inner-corner states,

respectively, for increasing loss contrasts from left to right. (e) and (f) Experimentally measured spatial power distributions of the outer-corner and inner-corner states, respectively, under three different dissipative loading conditions realized using polyurethane foams with different densities. The numerical and experimental results consistently show that spatial localizations of the topological states enhance obviously with the increasing loss contrast, namely, a tunable acoustic energy concentrator can be realized conveniently by change loss contrast.

This trend is more directly visualized from the full-wave acoustic-field distributions. Figs. 5(c) and (d) show the full-wave acoustic-field distributions of the outer-corner and inner-corner states, respectively, at three representative values of the loss contrast $\Delta\gamma$. At a relatively small value of $\Delta\gamma$, both modes already exhibit clear corner localization, although their fields still extend into neighboring lattice sites. As $\Delta\gamma$ increases, the outer-corner state becomes progressively concentrated at the external corners, whereas the inner-corner state contracts toward the corners of the central fractal cavity, accompanied by a gradual suppression of the field amplitude in the surrounding regions. These results show that the loss contrast not only determines the emergence of the topological phase, but also can tune the degree of spatial localization of both outer- and inner-corner states within the nontrivial regime.

To verify this loss-controlled localization experimentally, we further vary the dissipative loading of the fabricated sample and measure the corresponding spatial responses. Figs. 5(e) and 5(f) show the measured power distributions of the outer-corner and inner-corner states, respectively, under three different dissipative loading conditions. Three polyurethane foams with different densities (18 kg/m$^3$, 20 kg/m$^3$, and 25 kg/m$^3$ from left to right，respectively) are employed as absorptive materials to realize distinct levels of dissipation. With increasing dissipative loading, the outer-corner response becomes progressively concentrated at the external corners, while the inner-corner response exhibits increasingly pronounced localization around the internal corners. The measured evolution is consistent with the numerical trends shown in Figs. 5(c) and 5(d).

Fig. 5 therefore reveals the dual role of spatially nonuniform loss in this system. The loss contrast first drives the transition from the trivial phase to the nontrivial topological phase; once the system is in the nontrivial regime, further variation of $\Delta\gamma$ can continuously tune the spatial localization of the topological edge states without requiring any redesign of the fractal geometry or modification of the coupling configuration. This distinction between topology and localization identifies non-Hermitian loss not only as the driving parameter for the topological phase transition, but also as an independent degree of freedom for controlling the spatial confinement of hierarchical topological states.

## IV. Conclusion

In this work, we proposed a non-Hermitian route for realizing higher-order topological insulators in acoustic fractal lattices through theoretical analysis, numerical simulations, and experimental measurements. This extends the application boundaries of non-Hermitian topological physics into non-integer-dimensional systems and provides a novel paradigm for the multifunctional design of acoustic topological devices. The results show that merely modulating the spatial arrangement of physical loss is sufficient to effectively reshape the topological phases of the fractal acoustic system. Besides, the central role of loss contrast in determining the degree of acoustic energy localization within the system is revealed. Tuning non-Hermitian parameters by controlling the loss contrast not only provides a nascent mechanism to overcome the limitations of conventional Hermitian physics, but also enables the active and precise manipulation of extreme acoustic localization within complex fractal geometries. This physical capability to achieve highly concentrated acoustic energy on demand holds immense application potential for future practical engineering endeavors, including the development of highly sensitive acoustic detectors, efficient acoustic energy harvesters, and novel topological acoustic devices. Additionally, the non-Hermitian-fractal coupled topological mechanism revealed in this study can be directly extended to other classical wave systems, such as photonics and electronics，providing a crucial reference for the cross-disciplinary application of non-Hermitian topological devices.

## Acknowledgments

This work was supported by National Natural Science Foundation of China (Nos. 12572174, 12172183), the Natural Science Foundation of Zhejiang Province (No. LZ24A020001), International Science and Technology Cooperation Project lauched by Science and Technology Bureau of Ningbo City, Zhejiang Province, China (No. 2023H011). H. X. Wang is supported by the National Natural Science Foundation of China (Grant No. No. 12474432) and the Ningbo Yongjiang Talent Program (2024A-124-G).

## Appendix A. Tight-binding characterization of hierarchical topological states

To establish a direct correspondence with the full-wave results presented in Fig. 3, we further characterize the hierarchical topological states using the non-Hermitian tight-binding Hamiltonian under open edge conditions. This analysis directly relates the spectral positions of the eigenstates to their spatial distributions.

Fig. A1(a) shows the real part of the tight-binding spectrum of the G1 fractal lattice in the nontrivial regime. Four representative classes of eigenstates can be identified from their spatial profiles: bulk, outer-edge, outer-corner, and inner-corner states. The corner states lie near the center of the real-part spectral gap and are associated with the outer and internal fractal boundaries, respectively. Representative eigenstate distributions are shown in Figs. A1(c)–A1(e). State No. 65 is predominantly localized at the outermost corners of the lattice and is therefore identified as an outer-corner state. State No. 67 is concentrated around the corners of the central internal cavity and corresponds to an inner-corner state, whereas state No. 72 is mainly distributed along the external edge and is identified as an outer-edge state.

The corresponding spectrum of the G2 lattice is shown in Fig. A1(b). Owing to the additional level of recursive fractal construction, the number of internal boundaries increases substantially, giving rise to a richer set of localized modes. In addition to the outer-corner and outer-edge states, two types of inner-corner states and an inner-edge state can be distinguished. Their representative eigenstate distributions are displayed in Figs. A1(f)–A1(k). State No. 513 corresponds to an outer-corner state. States No. 516 and No. 507 represent two types of inner-corner states localized at geometrically distinct internal-edge corners. State No. 387 is primarily distributed along the external edge and is identified as an outer-edge state, whereas state No. 539 is mainly localized along an internal edge and corresponds to an inner-edge state. By contrast, state No. 213 remains extended throughout the lattice and is identified as a bulk state.

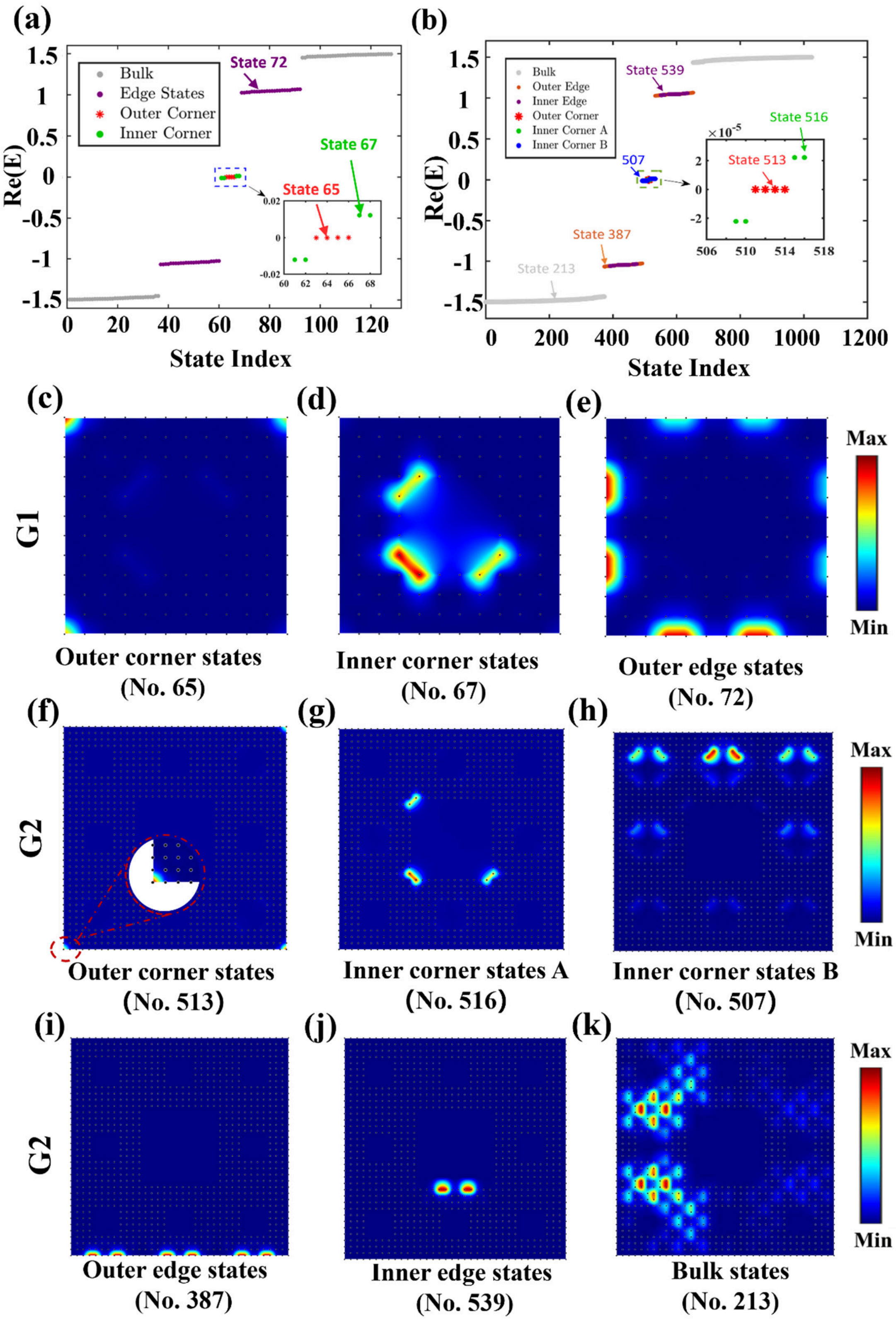

**FIG. A1.** Tight-binding characterization of hierarchical topological states in the G1 and G2 non-Hermitian fractal lattices. (a) Real part of the complex eigenvalue spectrum of the finite G1 lattice in the nontrivial regime. Bulk, outer-edge, outer-corner, and inner-corner states are distinguished

according to their spatial localization. The inset enlarges the corner-state manifold near the center of the gap. (b) Corresponding spectrum of the finite G2 lattice, exhibiting additional hierarchical inner-edge states. The inset shows an enlarged view of the corner-state manifold. (c)–(e) Spatial distributions of the representative G1 outer-corner (No. 65), inner-corner (No. 67), and outer-edge (No. 72) states, respectively. (f)–(k) Spatial distributions of the representative G2 outer-corner (No. 513), type-A inner-corner (No. 516), type-B inner-corner (No. 507), outer-edge (No. 387), inner-edge (No. 539), and bulk (No. 213) states, respectively.

These results reveal a direct connection between the recursive fractal geometry and the hierarchy of topological edge states. Importantly, Fig. A1 exhibits a clear correspondence with the full-wave results in Fig. 3 of the main text. The tight-binding model predicts not only the spectral separation of the edge states from the bulk states, but also their characteristic outer-corner, inner-corner, outer-edge, and inner-edge localization patterns. Although quantitative differences remain between the tight-binding and full-wave descriptions because of the finite dimensions and continuum geometry of the acoustic resonators, their spatial modal characteristics are consistent. This agreement indicates that the constructed non-Hermitian tight-binding Hamiltonian captures the essential physics underlying the hierarchical topological modes in the acoustic fractal lattice.

## Appendix B. Spectral response of a single resonator and extraction of loss parameters

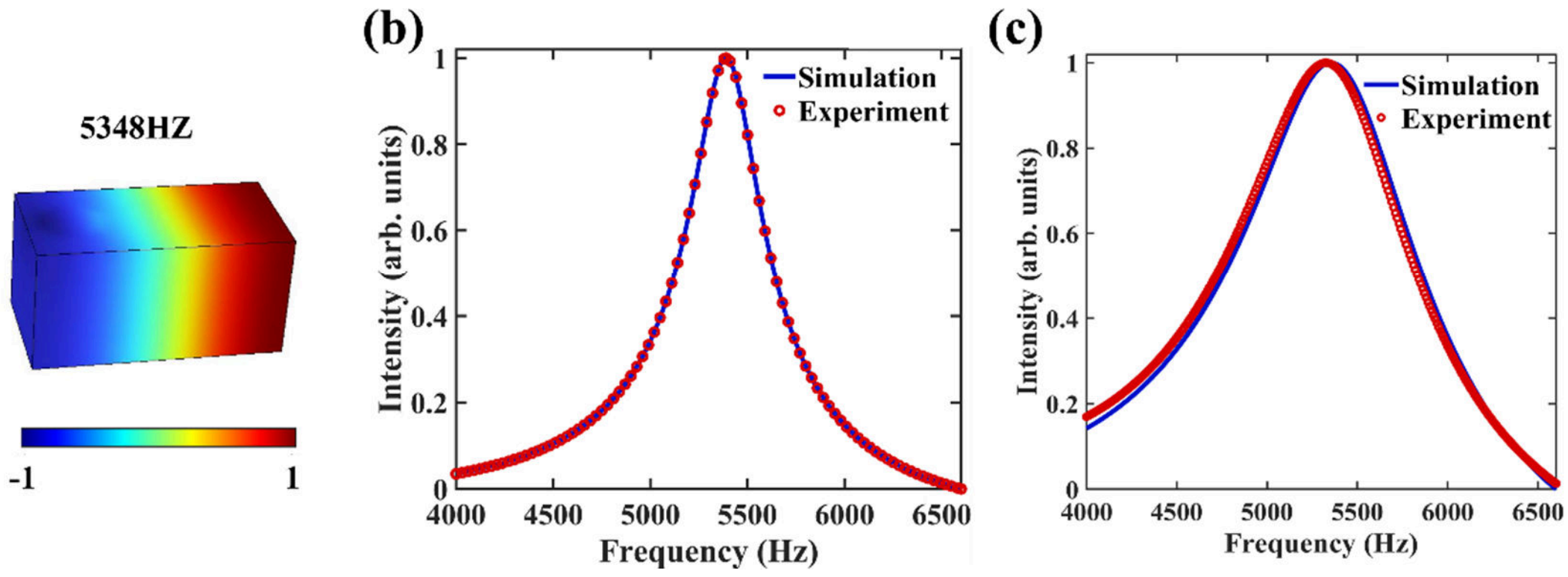

**FIG. B1.** Measured spectra of a single resonator. (a) Eigenmode of the dipole mode. (b), (c) Measured spectra of a single resonator with additional loss. Red dots represent the measured results, and the blue line indicates the numerical fitting results.

Rectangular acoustic resonant cavities were employed in the experiments. To accurately extract the simulation parameters, we conducted frequency response tests on a single cavity. As illustrated by the simulated field distribution in Fig. B1(a), the resonance frequency of interest is located around $\mathrm{f} \approx 5348$ Hz. This mode exhibits a distinct dipole distribution, which is conducive to forming strong coupling. For a single resonator without additional loss, the measured spectrum exhibits a peak at 5348 Hz, as shown in Fig. B1 (b). By fitting the experimental spectrum with the blue line, we can account for the background loss in the simulations by setting the speed of sound to $\mathrm{c} = 342.34 + 2.5\mathrm{i}\ \mathrm{m/s}$. The spectrum of the resonator with additional loss is also measured, as shown in Fig. B1 (c). The spectrum broadens, while the peak frequency exhibits almost no shift. Similarly, the speed of sound is estimated to be $\mathrm{c} = 342.34 + 20.8\mathrm{i}\ \mathrm{m/s}$. These values are subsequently used in the numerical simulations of the bulk dispersion and eigenmodes for the finite lattices.

## Appendix C. Calculation of the Hamiltonian for the fractal non-Hermitian acoustic system

For the finite non-Hermitian fractal lattice under open edge conditions, the tight-binding Hamiltonian can be written as:

$$H = \sum_{\langle m,n \rangle} t_{mn} \left( \hat{c}_m^\dagger \hat{c}_n + \hat{c}_n^\dagger \hat{c}_m \right) - i \sum_n \gamma_n \hat{c}_n^\dagger \hat{c}_m \qquad (1)$$

where $t_{mn}$ denotes the positive or negative nearest-neighbor coupling associated with the π-flux configuration. For each 16-site building block, additional loss is introduced at sites 2, 3, 5, 8, 9, 12, 14, and 15. Taking one 16-site building block as an example, the basis is arranged by first traversing the *x* direction and then the *y* direction as:

$$|\Psi\rangle = \left[ |1,1\rangle, |2,1\rangle, |3,1\rangle, |4,1\rangle \ldots, |4,4\rangle \right]^T \qquad (2)$$

Under this basis, the Hamiltonian of a single building block can be expressed as:

$$H = \begin{pmatrix} H_1 & V_1 & 0 & 0 \\ V_1 & H_2 & V_2 & 0 \\ 0 & V_2 & H_3 & V_1 \\ 0 & 0 & V_1 & H_4 \end{pmatrix} \tag{3}$$

where

$$H_1 = \begin{pmatrix} 0 & |k_1| & 0 & 0 \\ |k_1| & -i\Delta\gamma & |k_2| & 0 \\ 0 & |k_2| & -i\Delta\gamma & |k_1| \\ 0 & 0 & |k_1| & 0 \end{pmatrix}, \quad H_2 = \begin{pmatrix} -i\Delta\gamma & -|k_1| & 0 & 0 \\ -|k_1| & 0 & -|k_2| & 0 \\ 0 & -|k_2| & 0 & -|k_1| \\ 0 & 0 & -|k_1| & -i\Delta\gamma \end{pmatrix},$$

$$H_3 = \begin{pmatrix} -i\Delta\gamma & |k_1| & 0 & 0 \\ |k_1| & 0 & |k_2| & 0 \\ 0 & |k_2| & 0 & |k_1| \\ 0 & 0 & |k_1| & -i\Delta\gamma \end{pmatrix}, \quad H_4 = \begin{pmatrix} 0 & -|k_1| & 0 & 0 \\ -|k_1| & -i\Delta\gamma & -|k_2| & 0 \\ 0 & -|k_2| & -i\Delta\gamma & -|k_1| \\ 0 & 0 & -|k_1| & 0 \end{pmatrix} \tag{4}$$

$$V_1 = \begin{pmatrix} |k_1| & 0 & 0 & 0 \\ 0 & |k_1| & 0 & 0 \\ 0 & 0 & |k_1| & 0 \\ 0 & 0 & 0 & |k_1| \end{pmatrix}, \quad V_2 = \begin{pmatrix} |k_2| & 0 & 0 & 0 \\ 0 & |k_2| & 0 & 0 \\ 0 & 0 & |k_2| & 0 \\ 0 & 0 & 0 & |k_2| \end{pmatrix}.$$

Here, $k_1$ and $k_2$ denote the two alternating nearest-neighbor hopping amplitudes in the BBH-type hopping configuration. Their signs are arranged to generate an effective π flux. In the present model, the hopping magnitudes are chosen to be identical, $|k_1| = |k_2| = t$.

The Hamiltonian of the complete Sierpiński lattice is constructed by assembling these 16-site building blocks according to the fractal geometry. Neighboring retained building blocks are connected using the corresponding $k_2$ hopping terms following the same π-flux configuration. Under open edge conditions, couplings are included only between physically connected neighboring sites, while no coupling is introduced across the outer boundaries or across the removed regions of the Sierpiński lattice.

Since the Hamiltonian is non-Hermitian, the corresponding right and left eigenstates are obtained from:

$$H\left|\psi_n^R\right\rangle = E_n\left|\psi_n^R\right\rangle,$$

$$H^\dagger\left|\psi_n^L\right\rangle = E_n^*\left|\psi_n^L\right\rangle, \tag{5}$$

$$\left\langle\psi_m^L \mid \psi_n^R\right\rangle = \delta_{mn}.$$

where $E_n$ is the complex eigenvalue, and $|\psi_n^R\rangle$ and $|\psi_n^L\rangle$ denote the corresponding right and left eigenstates, respectively. The real part of $E_n$ characterizes the eigenfrequency component, whereas its imaginary part describes the modal loss. These biorthogonal eigenstates are subsequently used to calculate the non-Hermitian real-space quadrupole moment given in the main text.

**Appendix D**. **Details of the simulation and experimental setup**

1. Numerical simulations

All numerical simulations presented in this study were performed using the Pressure Acoustics Module in the commercial software COMSOL Multiphysics. In the numerical simulations, the air density of the background medium is set to 1.22 kg/m$^3$, and the real part of the speed of sound in air is set to 342.34 m/s. Loss is introduced by incorporating an imaginary term into the speed of sound. By adjusting the speed of sound to fit the measured and simulated spectra of a single resonator, $c$ is set to $342.34 + 2.5\mathrm{i}\ \mathrm{m/s}$ m/s for the resonator considering only background loss, whereas $\mathrm{c} = 342.34 + 20.8\mathrm{i}\ \mathrm{m/s}$ for the resonator with additional loss. Due to the large acoustic impedance mismatch between the photosensitive resin and air, the boundaries are modeled as sound-hard walls.

2. Experimental setup and measurement methods

2.1. Sample fabrication and experimental measurements

To experimentally verify the non-Hermitian-induced higher-order topological phases in the fractal system, we fabricated acoustic fractal lattice samples using the stereolithography 3D printing technology. The material used is a photosensitive resin (elastic modulus is 2765 MPa, density is 1.3g/cm$^3$) with a printing precision of ±0.1mm. The experimental sample comprises 128 acoustic resonators. Small holes with a diameter of 5 mm were drilled on both sides of each resonator, which can be opened or closed by two circular cover plates. In addition, 5 small holes with a diameter of 2 mm were drilled at both the left and right ends of the resonators corresponding to the blue rectangles in Fig. 4(a). During the experiment, acoustic waves generated by a loudspeaker are introduced into the sample through the small hole on one side of the resonator, and a microphone detects the signal through the small hole on the other side of the same resonator. This measurement is repeated for all resonators in the sample. During the measurements, the perforated cover plates at the excitation and detection sites are opened, while the remaining sites are sealed with cover plates. The blue resonators in the sample maintain an intact structure, whereas the drilled holes on the sidewalls of the red resonators are filled with blue sound-absorbing material (polyurethane foam) to achieve a differential loss configuration (Fig. 4(a)).

2.2. Experimental measurement system

The experimental system (Fig. D1) consists of a signal generator, a loudspeaker, the 3D-printed acoustic sample, a microphone, a conditioning amplifier, and a data acquisition system. The loudspeaker is driven by a signal generator (DG1022U, RIGOL) and a power amplifier (2706, B&K). The signal generator outputs a swept-frequency signal from 5100-5500Hz (with a resolution of 1 Hz), which is amplified by the power amplifier to drive the loudspeaker. The acoustic waves are incident through the small hole on one side of the resonator, and the transmitted signals are collected by a microphone (MPA416, BSWA) located at the small hole on the opposite side. The collected signals are then transmitted to a computer equipped with a control module (MC3242, BSWA) to detect the acoustic response of the waveguides.

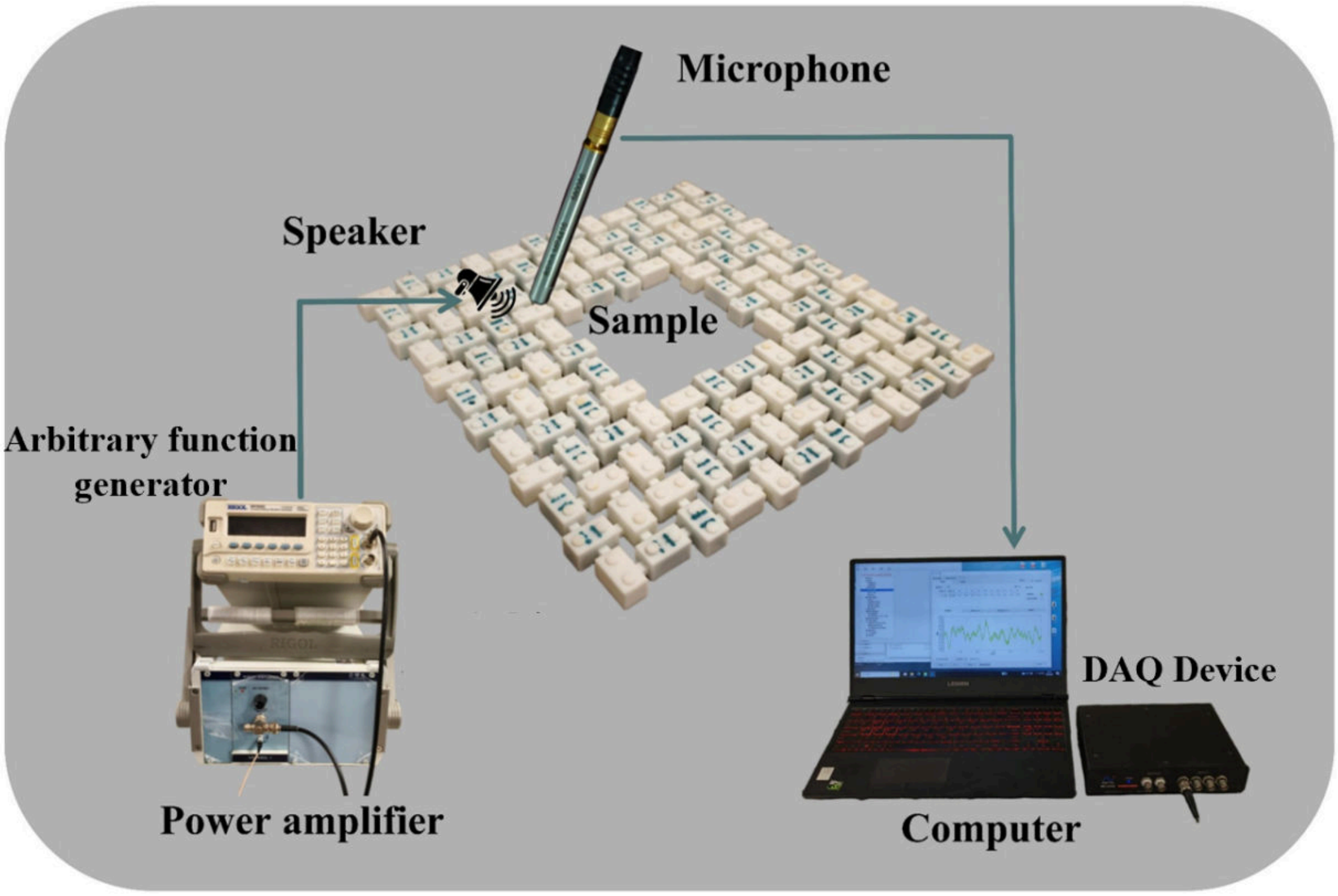

**FIG. D1.** Experimental setup

**References**:

[1] H. Xue, Y. Yang, and B. Zhang, Topological acoustics, Nature Reviews Materials **7**, 974 (2022).

[2] C. He, X. Ni, H. Ge, X.-C. Sun, Y.-B. Chen, M.-H. Lu, X.-P. Liu, and Y.-F. Chen, Acoustic topological insulator and robust one-way sound transport, Nature Physics **12**, 1124 (2016).

[3] J. Lu, C. Qiu, L. Ye, X. Fan, M. Ke, F. Zhang, and Z. Liu, Observation of topological valley transport of sound in sonic crystals, Nature Physics **13**, 369 (2017).

[4] Z. Yang, F. Gao, X. Shi, X. Lin, Z. Gao, Y. Chong, and B. Zhang, Topological Acoustics, Phys. Rev. Lett. **114**, 114301 (2015).

[5] W. A. Benalcazar, B. A. Bernevig, and T. L. Hughes, Quantized electric multipole insulators, Science **357**, 61 (2017).

[6] W. A. Benalcazar, B. A. Bernevig, and T. L. Hughes, Electric multipole moments, topological multipole moment pumping, and chiral hinge states in crystalline insulators, Phys. Rev. B **96**, 245115 (2017).

[7] B. Xie, H.-X. Wang, X. Zhang, P. Zhan, J.-H. Jiang, M. Lu, and Y. Chen, Higher-order band topology, Nature Reviews Physics **3**, 520 (2021).

[8] X. Ni, M. Weiner, A. Alù, and A. B. Khanikaev, Observation of higher-order

topological acoustic states protected by generalized chiral symmetry, Nature Materials **18**, 113 (2019).

[9] M. Serra-Garcia, V. Peri, R. Süsstrunk, O. R. Bilal, T. Larsen, L. G. Villanueva, and S. D. Huber, Observation of a phononic quadrupole topological insulator, Nature **555**, 342 (2018).

[10] H. Xue, Y. Yang, F. Gao, Y. Chong, and B. Zhang, Acoustic higher-order topological insulator on a kagome lattice, Nature Materials **18**, 108 (2019).

[11] X. Zhang, H.-X. Wang, Z.-K. Lin, Y. Tian, B. Xie, M.-H. Lu, Y.-F. Chen, and J.-H. Jiang, Second-order topology and multidimensional topological transitions in sonic crystals, Nature Physics **15**, 582 (2019).

[12] Q. Ma, Z. Pu, L. Ye, J. Lu, X. Huang, M. Ke, H. He, W. Deng, and Z. Liu, Observation of Higher-Order Nodal-Line Semimetal in Phononic Crystals, Phys. Rev. Lett. **132**, 066601 (2024).

[13] Y. Qi, Z. He, J. Li, B. Yan, Z. Zhang, and K. Deng, Experimental realization of non-Abelian higher-order topological insulators in two-dimensional phononic crystals, Phys. Rev. B **111**, 134113 (2025).

[14] Z. Wang, L. Ye, Z. Pu, Q. Ma, H. He, J. Lu, W. Deng, X. Huang, M. Ke, and Z. Liu, Higher-order topological transport protected by boundary Chern number in phononic crystals, Communications Physics **7**, 193 (2024).

[15] H. Xue, Y. Ge, H.-X. Sun, Q. Wang, D. Jia, Y.-J. Guan, S.-Q. Yuan, Y. Chong, and B. Zhang, Observation of an acoustic octupole topological insulator, Nature Communications **11**, 2442 (2020).

[16] E. J. Bergholtz, J. C. Budich, and F. K. Kunst, Exceptional topology of non-Hermitian systems, Rev. Mod. Phys. **93**, 015005 (2021).

[17] K. Ding, C. Fang, and G. Ma, Non-Hermitian topology and exceptional-point geometries, Nature Reviews Physics **4**, 745 (2022).

[18] K. Kawabata, K. Shiozaki, M. Ueda, and M. Sato, Symmetry and Topology in Non-Hermitian Physics, Phys. Rev. X **9**, 041015 (2019).

[19] H. Shen, B. Zhen, and L. Fu, Topological Band Theory for Non-Hermitian Hamiltonians, Phys. Rev. Lett. **120**, 146402 (2018).

[20] Q. Wu, S. Wang, H. Qian, Y. Wang, and G. Huang, Understanding of topological mode and skin mode morphing in 1D and 2D non-Hermitian resonance-based meta-lattices, Journal of the Mechanics and Physics of Solids **193**, 105907 (2024).

[21] Q. Wu, X. Xu, H. Qian, S. Wang, R. Zhu, Z. Yan, H. Ma, Y. Chen, and G. Huang,

Active metamaterials for realizing odd mass density, Proceedings of the National Academy of Sciences **120**, e2209829120 (2023).

[22] H. Qian, J. Chen, S. Wang, N. Shao, K. Liu, and G. Huang, Non-Hermitian Elastic Metabeams with Nonlinear and Nonlocal Coupling: Physical Realization and Topological Wave Manipulation, Advanced Functional Materials **36**, e18194 (2026).

[23] Y. Wang, J. Zhao, Q. Wu, X. Zhou, H. Jiang, W. Chen, M. Wang, and G. Huang, Physical Realization of an Anti-P-Pseudo-Hermitian Mechanical System, Phys. Rev. Lett. **136**, 096101 (2026).

[24] Y. Hu, J. Wu, P. Ye, W. Deng, J. Lu, X. Huang, Z. Wang, M. Ke, and Z. Liu, Acoustic Exceptional Line Semimetal, Phys. Rev. Lett. **134**, 116606 (2025).

[25] C. Shi, M. Dubois, Y. Chen, L. Cheng, H. Ramezani, Y. Wang, and X. Zhang, Accessing the exceptional points of parity-time symmetric acoustics, Nature Communications **7**, 11110 (2016).

[26] S. Yao and Z. Wang, Edge States and Topological Invariants of Non-Hermitian Systems, Phys. Rev. Lett. **121**, 086803 (2018).

[27] K. Yokomizo and S. Murakami, Non-Bloch Band Theory of Non-Hermitian Systems, Phys. Rev. Lett. **123**, 066404 (2019).

[28] L. Zhang et al., Acoustic non-Hermitian skin effect from twisted winding topology, Nature Communications **12**, 6297 (2021).

[29] B. Hu et al., Non-Hermitian topological whispering gallery, Nature **597**, 655 (2021).

[30] K. Kawabata, M. Sato, and K. Shiozaki, Higher-order non-Hermitian skin effect, Phys. Rev. B **102**, 205118 (2020).

[31] T. Liu, Y.-R. Zhang, Q. Ai, Z. Gong, K. Kawabata, M. Ueda, and F. Nori, Second-Order Topological Phases in Non-Hermitian Systems, Phys. Rev. Lett. **122**, 076801 (2019).

[32] X. Zhang, Y. Tian, J.-H. Jiang, M.-H. Lu, and Y.-F. Chen, Observation of higher-order non-Hermitian skin effect, Nature Communications **12**, 5377 (2021).

[33] J.-X. Zhong et al., Higher-order skin effect and its observation in an acoustic kagome lattice, Phys. Rev. B **111**, 014314 (2025).

[34] H. Fan, H. Gao, T. Liu, S. An, X. Kong, G. Xu, J. Zhu, C.-W. Qiu, and Z. Su, Reconfigurable topological modes in acoustic non-Hermitian crystals, Phys. Rev. B **107**, L201108 (2023).

[35] H. Gao, H. Xue, Z. Gu, T. Liu, J. Zhu, and B. Zhang, Non-Hermitian route to higher-order topology in an acoustic crystal, Nature Communications **12**, 1888 (2021).

[36] J. Wu, Y. Hu, Z. He, K. Deng, X. Huang, M. Ke, W. Deng, J. Lu, and Z. Liu, Hybrid-Order Skin Effect from Loss-Induced Nonreciprocity, Phys. Rev. Lett. **134**, 176601 (2025).

[37] K. Li, C. Wu, J. Kang, S. Liu, Q. Zhang, D. Zheng, W. Bao, Y. Zhou, and G. Chen, Emergence of higher-order topology in a loss-embedded kagome lattice, Phys. Rev. A **113**, 043523 (2026).

[38] T. Biesenthal, L. J. Maczewsky, Z. Yang, M. Kremer, M. Segev, A. Szameit, and M. Heinrich, Fractal photonic topological insulators, Science **376**, 1114 (2022).

[39] S. Pai and A. Prem, Topological states on fractal lattices, Phys. Rev. B **100**, 155135 (2019).

[40] Z. Yang, E. Lustig, Y. Lumer, and M. Segev, Photonic Floquet topological insulators in a fractal lattice, Light: Science & Applications **9**, 128 (2020).

[41] J. Li, Q. Mo, J.-H. Jiang, and Z. Yang, Higher-order topological phase in an acoustic fractal lattice, Science Bulletin **67**, 2040 (2022).

[42] S. Manna, S. Nandy, and B. Roy, Higher-order topological phases on fractal lattices, Phys. Rev. B **105**, L201301 (2022).

[43] S. Zheng, X. Man, Z.-L. Kong, Z.-K. Lin, G. Duan, N. Chen, D. Yu, J.-H. Jiang, and B. Xia, Observation of fractal higher-order topological states in acoustic metamaterials, Science Bulletin **67**, 2069 (2022).

[44] P. Lai, H. Liu, B. Xie, W. Deng, H. Wang, H. Cheng, Z. Liu, and S. Chen, Spin Chern insulator in a phononic fractal lattice, Phys. Rev. B **109**, L140104 (2024).

[45] J. Li, Y. Sun, Q. Mo, Z. Ruan, and Z. Yang, Fractality-induced topological phase squeezing and devil's staircase, Phys. Rev. Res. **5**, 023189 (2023).

[46] D. J. Salib, A. J. Mains, and B. Roy, Topological insulators on fractal lattices: A general principle of construction, Phys. Rev. B **110**, L241302 (2024).

[47] S. Manna and B. Roy, Inner skin effects on non-Hermitian topological fractals, Communications Physics **6**, 10 (2023).

[48] J. Sun, C.-A. Li, Q. Guo, W. Zhang, S. Feng, X. Zhang, H. Guo, and B. Trauzettel, Non-Hermitian quantum fractals, Phys. Rev. B **110**, L201103 (2024).

[49] W. Cao, X. Qin, W. Zhang, X. Zhou, and X. Zhang, Topological fractal braiding of non-Hermitian bands, Communications Physics **8**, 391 (2025).